\pdfoutput=1 
\documentclass{article}%
\usepackage{amssymb}
\usepackage{amsfonts}
\usepackage{amsmath}
\usepackage{graphicx}%
\begin{document}

\title{Phenomenological mass relation for free massive stable particles and
estimations of neutrino and graviton masses }
\author{Dimitar Valev\\\textit{Stara Zagora Department, Solar-Terrestrial Influences Laboratory,}\\\textit{\ Bulgarian Academy of Sciences, 6000 Stara Zagora, Bulgaria}}
\maketitle

\begin{abstract}
The ratio between the proton and electron masses was shown to be close to the
ratio between the shortest lifetimes of particles, decaying by the
electromagnetic and strong interactions. The inherent property of each
fundamental interaction is defined, namely the Minimal lifetime of the
interaction (\textit{MLTI}). The rest mass of the Lightest free massive stable
particle (\textit{LFMSP}), acted upon by a particular interaction, is shown to
be inversely proportional to \textit{MLTI}. The found mass relation unifies
the masses of four stable particles of completely different kinds (proton,
electron, electron neutrino and graviton) and covers an extremely wide range
of values, exceeding 40 orders of magnitude. On the basis of this mass
relation, the electron neutrino and graviton masses have been approximately
estimated to $6.5\times10^{-4}$ $%
\operatorname{eV}%
$ and $\hbar H/c^{2}\approx1.5\times10^{-33}$ $%
\operatorname{eV}%
$, respectively. Besides, the last value has been obtained independently by
dimensional analysis by means of three fundamental constants, namely the speed
of light in vacuum ($c$), reduced Planck constant ($\hbar$) and Hubble
constant ($H$). It was shown that the rest energy of \textit{LFMSP}, acted
upon by a particular interaction, is close to Breit-Wigner's energy width of
the shortest living state, decaying by the respective interaction.


\end{abstract}

\section{Introduction}

Although the neutrino and the graviton belong to different particle kinds
(neutral lepton and quantum of the gravitation, respectively), they have some
similar properties. Both particles are not acted upon by the strong and the
electromagnetic interactions, which makes their detection and investigation
exceptionally difficult. Besides, both have masses that are many orders of
magnitude lighter than the masses of the rest particles and they are generally
accepted to be massless.

Decades after the experimental detection of the neutrino \cite{Reines}, it was
generally accepted that the neutrino rest mass $m_{0\nu}$ is rigorously zero.
The first experiment, hinting that the neutrino probably possesses a mass, is
dated back to the $60-$ies \cite{Davis}. The total flux of neutrinos from the
Sun is about $3$ times lower than the one, predicted by theoretical solar
models. This discrepancy can be explained if some of the electron neutrinos
transform into another neutrino flavor. Later, the experimental observations
showed that the ratio between the atmospheric $\nu_{\mu}\ $and $\nu_{e}%
\ $fluxes was less than the theoretical predictions \cite{Hirata, Casper}.
Again the discrepancy could be explained by the neutrino oscillations. The
crucial experiments with the $50\ $Kton neutrino detector Super-Kamiokande
found strong evidence for oscillations (and hence - mass) in the atmospheric
neutrinos \cite{Fukuda}.

The direct neutrino measurements allow to limit the neutrino mass. The upper
limit for the mass of the lightest neutrino flavor $\nu_{e}\ $was obtained
from experiments for measurement of the high-energy part of the tritium
$\beta-\ $spectrum and recent experiments yield $2%
\operatorname{eV}%
$ upper limit \cite{Weinheimer, Lobashev}. As a result of the recent
experiments, the upper mass limits of $\nu_{\mu}\ $and $\nu_{\tau}\ $were
found to be $170\ $K$%
\operatorname{eV}%
$ \cite{Assamagan} and $18.2\
\operatorname{MeV}%
$ \cite{Barate}, respectively. The Solar and atmospheric neutrino experiments
allow to find the square mass differences $\Delta m_{12}^{2}=m_{2}^{2}%
-m_{1}^{2}$ and $\Delta m_{23}^{2}=m_{3}^{2}-m_{2}^{2}$, but not the absolute
values of the neutrino masses. The astrophysical constraint of the neutrino
mass is $%
{\textstyle\sum}
m_{\nu}<\ 2.2\
\operatorname{eV}%
$ \cite{Bahcall}. The recent extensions of the Standard model lead to non-zero
neutrino masses, which are within the large range of $10^{-6}%
\operatorname{eV}%
\div10%
\operatorname{eV}%
$.

Similarly to the case with the neutrino before $1998$, the prevailing current
opinion is that the quantum of the gravitation (graviton) is massless. This
opinion is connected with Einstein's theory of General Relativity, where the
gravitation is described by a massless field of spin $2 $ in a generally
covariance manner. The nonzero graviton mass leads to a finite gravitation
range $r_{g}\sim\lambda_{g}/2\pi=\hbar/(m_{g}c),$ where $\lambda_{g}\ $is
Compton wavelength of the graviton. The lowest astrophysical limit of the
graviton mass is obtained by rich galactic clusters $m_{g}<2\times
10^{-29}h^{-1}%
\operatorname{eV}%
$ \cite{Goldhaber}, where $h\approx0.70$ is a dimensionless Hubble constant.
In this case no difference was observed between Yukawa's potential for the
massive graviton and Newton's potential for the massless graviton.

It has been obtained a value of the graviton mass $m_{g}\sim4.3\times10^{-34}%
\operatorname{eV}%
$ for an infinite stationary universe \cite{Woodward}, although the expansion
of the Universe is a fact, long ago established. The mass $m_{i\min}\ $of the
Lightest free massive stable particle (\textit{LFMSP}), acted upon by a
particular interaction, is shown to be proportional to the coupling constant
of the respective interaction at extremely low energy \cite{Valev a}. The
graviton and electron neutrino masses have been estimated by this approach to
$m_{g}\sim2.3\times10^{-34}%
\operatorname{eV}%
$ and $m_{\nu e}\sim2.1\times10^{-4}%
\operatorname{eV}%
$, respectively.

\section{Minimal lifetime of a fundamental interaction and a mass relation for
free massive stable particles}

Among the multitude of particles, several free particles are notable, which
are stable or at least their lifetimes are longer than the age of the Universe
-- the proton ($p$), electron ($e$), neutrino ($\nu$) (three flavors),
graviton ($g$) and photon ($\gamma$). Only \textit{free massive stable}
particles are examined in this paper. Quarks and gluons are bound in hadrons
by confinement and they cannot be immediately detected in the experiments, and
the photon is massless. Therefore, these particles are not a subject of this paper.

A measure for the interaction strength is a dimensionless quantity - the
coupling constant of the interaction ($\alpha_{i}$), which is determined from
the cross section of the respective processes. Generally, it is known that the
bigger the strength (coupling constant) of an interaction, the quicker (with
shorter duration $\tau$) are the processes, ruled by this interaction.
Actually, the typical lifetime of resonances, decaying by the strong
interaction ($\tau_{s}$) is from $10^{-24}\
\operatorname{s}%
$ to $10^{-23}\
\operatorname{s}%
$, the time of the radiative decay ($\tau_{e}$) of particles and excited
stages of nuclei is from $10^{-21}\
\operatorname{s}%
$ to $10^{-12}\
\operatorname{s}%
$ and, the lifetime ($\tau_{w}$) of particles decaying by the weak interaction
is from $10^{-12}\
\operatorname{s}%
$ to $10^{3}\
\operatorname{s}%
$. Since \textquotedblleft the age of the Universe\textquotedblright\ is
$H^{-1}\sim4.3\times10^{17}%
\operatorname{s}%
\approx1.37\times10^{10}$ years, the lifetime of particles decaying by the
gravitational interaction is $\tau_{g}\gtrsim H^{-1}$.

The fastest process by the strong interaction is the $f_{0}(400-1200)$
resonance decay, having $\tau_{s\min}\approx8\times10^{-25}%
\operatorname{s}%
$ \cite{Groom}. The fastest process by the electromagnetic interaction is the
radiative decay of the super hot nuclei $\tau_{e\min}\sim\lambda_{e}/(2\pi
c)\approx1.3\times10^{-21}%
\operatorname{s}%
$, where $\lambda_{e}\ $is Compton wavelength of the electron
\cite{Karnaukhov}. The fastest decay by the weak interaction is flavor
transformation of the bottom and charmed quarks with $\tau_{w\min}\sim10^{-12}%
\operatorname{s}%
$. The minimal lifetime $\tau_{g\min}\ $of particles, decaying by the
gravitational interaction is unknown, therefore we suggest that $\tau_{g\min
}\sim H^{-1}\approx4.3\times10^{17}%
\operatorname{s}%
$. Thus, the cosmological expansion of the Universe is considered a
manifestation of the gravitational decay. Therefore, the minimal lifetime
($\tau_{i\min}$) of particles, decaying by a particular interaction, appears a
\textit{unique inherent property} of each interaction, below named Minimal
lifetime of the interaction (\textit{MLTI}).

The ratio between the proton and electron masses is $m_{p}/m_{e}\approx1836$.
On the other side, the ratio between the minimal lifetimes of electromagnetic
and strong interactions is $\tau_{e\min}/\tau_{s\min}\sim1625$. The two ratios
differentiate by less than $13\ \%$. Therefore, the ratio between the proton
and electron masses is close to the ratio between the minimal lifetimes of the
electromagnetic and strong interactions:%

\begin{equation}
\frac{m_{p}}{m_{e}}\sim\frac{\tau_{e\min}}{\tau_{s\min}} \label{eqn1}%
\end{equation}

The proton and electron are the Lightest free massive stable particles
(\textit{LFMSP}), acted upon by the strong and electromagnetic interactions,
respectively. This relation is remarkable since it connects the masses of
\textit{LFMSP}, acted upon by the strong and electromagnetic interactions and
the respective \textit{MLTI}. The relation (\ref{eqn1}) suggests that the mass
of \textit{LFMSP}, acted upon by the strong (or the electromagnetic)
interaction is inversely proportional to the respective \textit{MLTI}, i.e.
$m_{p}\approx k/\tau_{s\min}\ $and $m_{e}\approx k/\tau_{e\min}$, where $k$ is
a constant. Therefore, it is interesting to examine whether this rule will be
valid both for the weak interaction, whose \textit{MLTI} is several orders of
magnitude longer than the minimal lifetime of the electromagnetic interaction
and for the gravity, whose minimal lifetime is dozens orders of magnitude
longer than minimal lifetime of the weak interaction. \textit{LFMSP} acted
upon by the weak interaction is the electron neutrino and \textit{LFMSP} acted
upon by the gravity most probably appears the hypothetical graviton. Although
the rest masses of the two particles are still unknown, the direct neutrino
mass experiments and the theoretical models suggest that the $\nu_{e}\ $mass
is between $10^{-6}%
\operatorname{eV}%
$ and $2%
\operatorname{eV}%
$, i.e. $\nu_{e}\ $is several orders of magnitude lighter than the electron.
Again, the astrophysical constraints allow to find the upper limits of the
graviton mass and according to these constraints, if the graviton really
exists, its mass would be less than $3\times10^{-29}%
\operatorname{eV}%
$, i.e. dozens orders of magnitude lighter than $\nu_{e}$. Table \ref{Table 1}
presents \textit{MLTI}, as well as the masses of \textit{LFMSP}, acted upon by
the respective interaction. The experimental upper limits of the electron
neutrino and graviton masses are also presented. Table \ref{Table 1} shows
that the mass of \textit{LFMSP} acted upon by a particular interaction
decreases with the increase of \textit{MLTI}.%

\begin{table}[htb] \centering
\caption{MLTI and the masses of LFMSP acted upon by the respective
interaction.}%
\begin{tabular}
[c]{|l|l|l|l|}\hline
Interaction & \textit{MLTI} $(%
\operatorname{s}%
)$ & \textit{LFMSP} & Experimental mass or\\
&  &  & mass limit of \textit{LFMSP} $(%
\operatorname{eV}%
)$\\\hline
Strong & $8\times10^{-25}$ & $p$ & $9.38\times10^{8}$\\\hline
Electromagnetic & $1.3\times10^{-21}$ & $e$ & $5.11\times10^{5}$\\\hline
Weak & $10^{-12}$ & $\nu_{e}$ & $0<m<2$\\\hline
Gravitational & $4.3\times10^{17}$ & $g$ & $<3\times10^{-29}$\\\hline
\end{tabular}
\label{Table 1}%
\end{table}%

\begin{figure}
[t]
\begin{center}
\includegraphics[
natheight=5.166400in,
natwidth=8.507200in,
height=2.9871in,
width=4.9009in
]%
{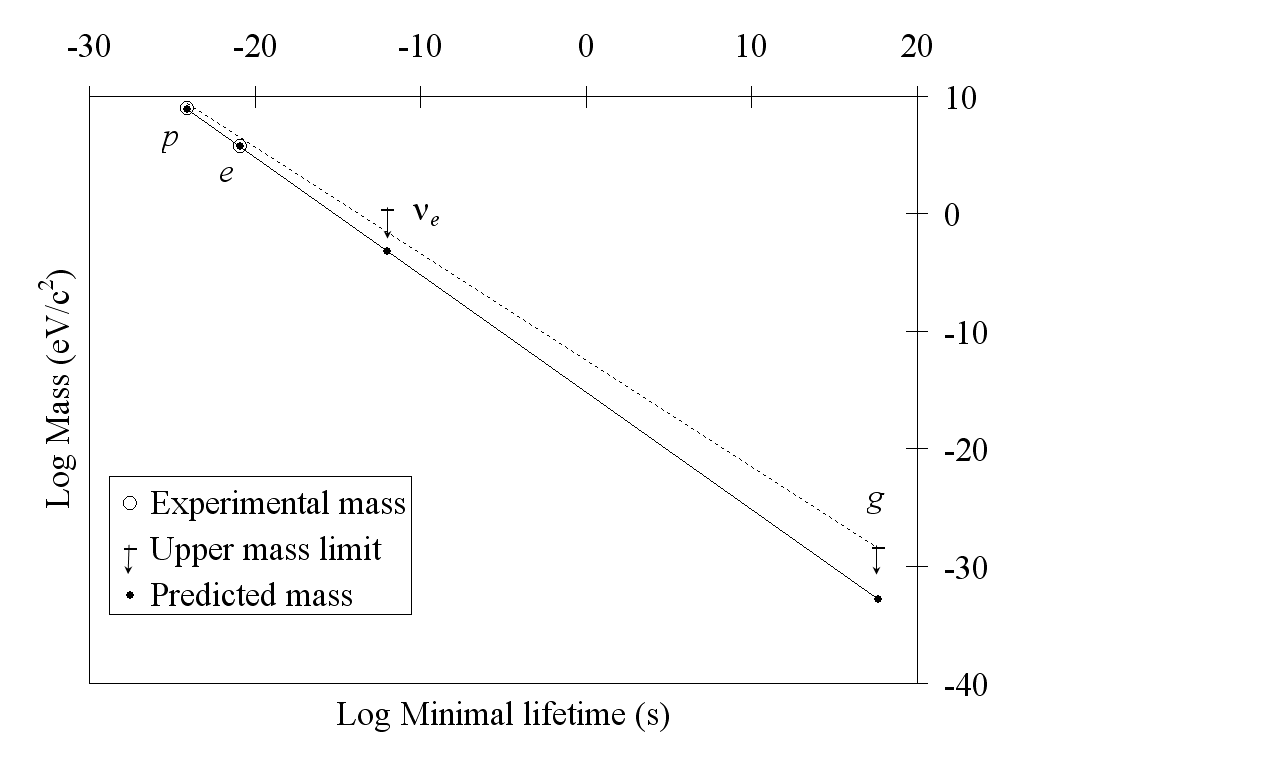}%
\caption[Mass Relation]{Dependence between the mass of \textit{LFMSP} acted
upon by a particular interaction and \textit{MLTI}. The dashed line represents
the approximation (2) of $e,\ p$ and the upper limit masses of $g$ and
$\nu_{e}$. The solid line represents the strict inverse linear approximation
($S=-1$).}%
\label{Figure1}%
\end{center}
\end{figure}

The data from Table \ref{Table 1} have been presented in a double-logarithmic
scale in Fig. \ref{Figure1}, which shows that the trend is clearly expressed.%

The points in Fig. \ref{Figure1}, corresponding to the electron and proton
masses and to the upper limit masses of the electron neutrino and graviton,
are approximated by the least squares with a power law:%

\begin{equation}
\log m_{\min}=-0.90\log\tau_{\min}-12.20 \label{eqn2}%
\end{equation}

Although this approximation is only on four points, the found correlation is
close and the correlation coefficient reaches $r=0.998$, which supports the
power law. The modulus of the slope ($S$) is little smaller than one and that
is why it can be said that the regression is close to a linear one. In
addition, it should be reminded that instead of the electron neutrino and
graviton masses, their upper limit values are used, which produce a certain
underestimation of the $S$ value. This approximation shows that the mass of
\textit{LFMSP}, acted upon by a particular interaction, increases with the
decrease of the respective \textit{MLTI} by a power law with $S\sim-1$, i.e.
close to the inverse linear one. The inverse proportionality of the proton
mass to the strong coupling constant and of the electron mass to the fine
structure constant also support inverse linear dependence (\textit{without
intercept}). Thus, the experimental data suggest inverse linear dependence
($S=-1$) between the mass of \textit{LFMSP} acted upon by a particular
interaction and \textit{MLTI}:%

\begin{equation}
\log m_{\min}=-\log\tau_{\min}-k_{0} \label{eqn3}%
\end{equation}

where $k_{0}$ is a constant.

The expression (\ref{eqn3}) transforms into:%

\begin{equation}
m_{\min}\tau_{\min}=10^{-k_{0}}=k \label{eqn4}%
\end{equation}

In this way the experimental data and constraints suggest that the mass of
\textit{LFMSP}, acted upon by a particular interaction, is inversely
proportional to the respective \textit{MLTI}:%

\begin{equation}
m_{i\min}=\frac{k}{\tau_{i\min}} \label{eqn5}%
\end{equation}

where $k=m_{e}\tau_{e\min}=6.54\times10^{-16}%
\operatorname{eV}%
\operatorname{s}%
=1.16\times10^{-51}%
\operatorname{kg}%
\operatorname{s}%
$ is a constant, $i=1,2,3,4$ -- index for each interaction and \textit{LFMSP}
acted upon by the respective interaction.

In consideration of $\tau_{e\min}\sim\lambda_{e}/(2\pi c)$ and Compton formula
$\lambda_{e}=h/(m_{e}c)$, the mass relation (\ref{eqn5}) would be transformed
in the equivalent mass formula:%

\begin{equation}
m_{i\min}=\frac{m_{e}\tau_{e\min}}{\tau_{i\min}}\sim\frac{\hbar}{c^{2}%
\tau_{i\min}} \label{eqn6}%
\end{equation}

\section{Neutrino and graviton mass estimations}

The found mass relation (\ref{eqn5}), and equivalent mass formula
(\ref{eqn6}), could be examined by the strong interaction because the proton
mass is measured with high precision. The application of the mass relation on
the strong interaction predicts the lightest stable hadron mass $m_{p}%
\approx819%
\operatorname{MeV}%
$. Thus, the proton mass value obtained by the mass relation (\ref{eqn5}) is
only $12.7\ \%$ lower than the experimental value of $m_{p}$. This result
confirms the reliability of the found mass relation and shows that this
relation possesses heuristic power. The application of the mass relation
(\ref{eqn5}) on the weak interaction allows to evaluate the mass of the
electron neutrino $m_{\nu e}\approx6.5\times10^{-4}%
\operatorname{eV}%
$. This value is in order of magnitude of the estimation of the electron
neutrino mass, found in \cite{Valev a}.

The above obtained value $m_{\nu e}\approx6.5\times10^{-4}%
\operatorname{eV}%
$ and the results from the solar and atmospheric neutrino experiments allow to
estimate the masses of the heavier neutrino flavor $-$\ $\nu_{\mu}$ and
$\nu_{\tau}$. The results from the Super Kamiokande experiment lead to square
mass difference $\Delta m_{23}^{2}\sim2.7\times10^{-3}%
\operatorname{eV}%
^{2}$ \cite{Gonzalez-Garcia}. Recent results on solar neutrinos provide hints
that the Large mixing angle (\textit{LMA}) of Mikheyev-Smirnov-Wolfenstein
(\textit{MSW}) solution is more probable than the Small mixing angle
(\textit{SMA}) \cite{Krastev}. The \textit{LMA} leads to $\Delta m_{12}%
^{2}\sim7\times10^{-5}%
\operatorname{eV}%
^{2}$ \cite{Maltoni} and the \textit{SMA} leads to $\Delta m_{12}^{2}%
\sim6\times10^{-6}%
\operatorname{eV}%
^{2}$ \cite{Albright}. In this way both solutions yield $m_{\nu\tau}\sim0.05%
\operatorname{eV}%
$. The most appropriate \textit{LMA} yields $m_{\nu\mu}\sim8.4\times10^{-3}%
\operatorname{eV}%
$ and the \textit{SMA} leads to $m_{\nu\mu}\sim2.5\times10^{-3}%
\operatorname{eV}%
$. Thus, the obtained values of the neutrino masses support the normal
hierarchy case. These values are close to the predictions of the simple
$SO(10)$ model for the neutrino masses \cite{Dermisek}.

In consideration of $\tau_{g}\sim H^{-1}$, the mass formula (\ref{eqn6})
allows to estimate the graviton mass:%

\begin{equation}
m_{g}\sim\frac{\hbar}{c^{2}\tau_{g}}\sim\frac{\hbar H}{c^{2}}\approx
1.5\times10^{-33}%
\operatorname{eV}
\label{eqn7}%
\end{equation}

The predicted masses of four \textit{LFMSP} are presented in Table
\ref{Table 2}, where it can be seen that the fitting of the predicted values
and the experimental data is satisfactory.%

\begin{table}[htb] \centering
\caption{Experimental and predicted values of the masses of LFMSP.}%
\begin{tabular}
[c]{|l|l|l|l|}\hline
{\normalsize Particle} & {\normalsize Experimental mass} &
{\normalsize Predicted mass }$(%
\operatorname{eV}%
)$ & Predicted mass $(%
\operatorname{eV}%
)$\\
& or mass limit $(%
\operatorname{eV}%
)$ & [This paper] & \cite{Valev a}\\\hline
$p$ & $9.38\times10^{8}$ & $8.19\times10^{8}$ & $9.8\times10^{8}$\\\hline
$e$ & $5.11\times10^{5}$ & $5.11\times10^{5}$ & $5.11\times10^{5}$\\\hline
$\nu_{e}$ & $0<m<2$ & $6.5\times10^{-4}$ & $2.1\times10^{-4}$\\\hline
$g$ & $<3\times10^{-29}$ & $1.5\times10^{-33}$ & $2.3\times10^{-34}$\\\hline
\end{tabular}
\label{Table 2}%
\end{table}%

The exceptionally small graviton rest mass seriously impedes its experimental
determination. Yet, it can be expected that appropriate astrophysical or
laboratory experiments would be conducted for this aim. Probably, the
investigations of the large-scale structure of the universe and the microwave
background radiation would contribute to the astrophysical estimation of the
graviton mass. The massive graviton might turn of considerable importance for
the description of the processes in the nuclei of the active galaxies and
quasars, the gravitational collapse as well as for the improvement of the
cosmological models.

Besides, the formula (\ref{eqn7}) for graviton mass could be obtained
\textit{independently} by dimensional analysis. Actually, by means of three
fundamental constants, namely the speed of light in vacuum ($c$), reduced
Planck constant ($\hbar$) and Hubble constant ($H$), a mass dimension quantity
$m_{x}\ $could be constructed:%

\begin{equation}
m_{x}=kc^{\alpha}\hbar^{\beta}H^{\gamma} \label{eqn8}%
\end{equation}

where $k$ is dimensionless parameter of the order of magnitude of one and
$\alpha,\beta$ and $\gamma$ are unknown exponents, which will be determined by
dimensional analysis below.

Dimensional analysis has been successfully used in \cite{Valev b} for
estimation of mass of the observable universe by means of following
fundamental constants -- the speed of light ($c$), universal gravitational
constant ($G$) and Hubble constant ($H$).

The dimensions of the left and right sides of the equation (\ref{eqn8}) must
be equal:%

\begin{equation}
\lbrack m_{x}]=[c]^{\alpha}[\hbar]^{\beta}[H]^{\gamma} \label{eqn9}%
\end{equation}

Taking into account the dimensions of quantities in formula (\ref{eqn9}) we obtain:%

\begin{equation}
L^{0}T^{0}M^{1}=(LT^{-1})^{\alpha}(ML^{2}T^{-1})^{\beta}(T^{-1})^{\gamma
}=L^{\alpha+2\beta}T^{-\alpha-\beta-\gamma}M^{\beta} \label{eqn10}%
\end{equation}

where $L,T$ and $M$ are dimensions of length, distance and mass, respectively.

From (\ref{eqn10}) we obtain the system of linear equation:%

\begin{align}
\alpha+2\beta &  =0\nonumber\\
-\alpha-\beta-\gamma &  =0\label{eqn11}\\
\beta &  =1\nonumber
\end{align}

Solving the system we find the exponents $\alpha=-2,\beta=1,\gamma=1$.
Replacing the obtained values of the exponents in equation (\ref{eqn8}) we
find the formula (\ref{eqn12}) for the graviton mass:%

\begin{equation}
m_{x}\sim\frac{\hbar H}{c^{2}} \label{eqn12}%
\end{equation}

Although this formula has been found by totally different approach, it
coincides with formula (\ref{eqn7}), which reinforces the found
phenomenological mass relation (\ref{eqn6}).

According Big Bang cosmology \cite{Hubble}, Hubble constant decrease with age
of the universe, therefore the found graviton mass (\ref{eqn7}) slowly
decrease with time. On the other hand, according Tired Light model
\cite{Zwicky} and Steady State theory \cite{Bondi}, the Hubble constant $H$ is
truly a constant not only in all directions, but at all time. Therefore, the
graviton mass is truly constant in the framework of Tired Light model and
Steady State theory.

\section{Discussions.}

From mass formula (\ref{eqn6}) we obtain:%

\begin{equation}
m_{i\min}c^{2}\sim\frac{\hbar}{\tau_{i\min}}\approx\Gamma_{i\min}
\label{eqn13}%
\end{equation}

where $\Gamma_{i\min}$ is Breit-Wigner's energy width of the shortest living
state, decaying by the respective interaction.

Therefore, the rest energy of \textit{LFMSP} acted upon by a particular
interaction is close to Breit-Wigner's energy width of the shortest living
state, decaying by the respective interaction. It should be reminded that here
$\tau_{i\min}\ $isn't the lifetime of the particle (it is stable) but
$\tau_{i\min}\ $is the respective \textit{MLTI}.

The mass formula (\ref{eqn6}) would be written in the form:%

\begin{equation}
m_{i\min}c^{2}\tau_{i\min}\sim\hbar\label{eqn14}%
\end{equation}

The mass ($m_{i}$) of each free massive stable particle, acted upon by a
particular interaction $m_{i}\geq m_{i\min}\ $and the lifetime ($\tau_{i}$) of
particles decaying by the respective interaction $\tau_{i}\geq\tau_{i\min}$.
As a result, the inequality (\ref{eqn15}) is obtained:%

\begin{equation}
m_{i}c^{2}\tau_{i}\geq\hbar\label{eqn15}%
\end{equation}

The comparison of (\ref{eqn15}) and the Uncertainty Principle $\Delta E\Delta
t=\Delta(mc^{2})\Delta t\geq\hbar\ $shows that the inequality (\ref{eqn15}),
which results from the mass formula (\ref{eqn6}), is related to the
Uncertainty Principle. Thus, equation (\ref{eqn14}) appears a boundary case of
the Uncertainty Principle at minimal allowed values of the rest energy and
lifetime of the real particles. In this case, however, a more general
interpretation of the Uncertainty Principle will be necessary since equation
(\ref{eqn14}) relates the mass ($m_{i\min}$) of \textit{LFMSP} acted upon by a
particular interaction with minimal lifetime ($\tau_{i\min}$) of particles
(states), decaying by the respective interaction. The future complete Unified
theory of the four interactions would give theoretical explanation of this
dependence. Most probably $\tau_{i\min}\ $determines inability of the
respective interaction to create free massive stable particles, possessing
rest mass $m_{i}\leq m_{i\min}=\hbar/(c^{2}\tau_{i\min})$. In other words, the
stronger an interaction the smaller is \textit{MLTI} and the heavier is
\textit{LFMSP}, which it is capable to create.

The mass relation (\ref{eqn16}) has been obtained in \cite{Valev a} by a
similar phenomenological approach:%

\begin{equation}
m_{i\min}=\frac{m_{e}}{\alpha}\alpha_{i}(0) \label{eqn16}%
\end{equation}

where $m_{i\min}\ $is the mass of \textit{LFMSP}, acted upon by a particular
interaction, $\alpha_{i}(0)\ $is the coupling constant of a particular
interaction at extremely low energy $E\sim m_{e}c^{2}\ $and $\alpha$ is the
fine structure constant.

From (\ref{eqn16}) and (\ref{eqn6}) we obtain:%

\begin{equation}
\alpha_{i}(0)=\frac{\alpha m_{i\min}}{m_{e}}=\frac{\hbar\alpha}{m_{e}c^{2}%
\tau_{i\min}}=\frac{\alpha\lambda_{e}/2\pi}{c\tau_{i\min}}\sim\alpha\frac
{\tau_{e\min}}{\tau_{i\min}}\approx\frac{\tau_{n}}{\tau_{i\min}} \label{eqn17}%
\end{equation}

where $\tau_{n}\ $is the nuclear time.

Equation (\ref{eqn17}) supports the natural suggestion that the coupling
constant of a particular interaction at extremely low energy $\alpha_{i}%
(0)\ $is inversely proportional to \textit{MLTI} and it determines from the
ratio $\tau_{n}/\tau_{i\min}$.

It is worth noting that each \textit{LFMSP}, acted upon by a particular
interaction also appears the lightest free massive particle, possessing the
respective \textit{universal conserving quantity} - baryon number, electric
charge, lepton number and mass. Actually, $p,e,\nu_{e}$, and $g$ are
\textit{LFMSP}, acted upon by the strong, electromagnetic, weak and
gravitational interaction, respectively, and they also appear lightest free
massive particles, possessing baryon number, electric charge, lepton number
and mass. It should also be mentioned that all four lightest free massive
particles, possessing universal conserving quantity are stable or, at least
their lifetimes are longer than the age of the Universe.

The massive graviton rises severe challenges before the modern unified
theories. Among them are van Dam-Veltman-Zakharov ($vDVZ$) discontinuity
\cite{van Dam, Zakharov} and the violation of the gauge invariance and the
general covariance. There are, however, already encouraging attempts to solve
$vDVZ$ discontinuity in anti de Sitter ($AdS$) background \cite{Dvali, Kogan}.

\section{Conclusions}

It was found that the ratio between the proton and electron masses is close to
the ratio between the shortest lifetime of particles, decaying by the
electromagnetic and strong interaction $m_{p}/m_{e}\sim\tau_{e\min}%
/\tau_{s\min}$. The inherent property of each fundamental interaction is
defined, namely the Minimal lifetime of the interaction (\textit{MLTI}).
Inverse proportionality has been found between \textit{MLTI} as well as the
rest mass of the \textit{LFMSP} acted upon by the respective interaction
$m_{i\min}=k/\tau_{i\min}\sim\hbar/(c^{2}\tau_{i\min})$.

The rest mass of the electron neutrino has been obtained by this approach to
$m_{\nu e}\approx6.5\times10^{-4}%
\operatorname{eV}%
$. The masses of the heavier neutrino flavors have been estimated by the
results of the solar and atmospheric neutrino experiments. The mass of
$\nu_{\tau}\ $is estimated to $0.05%
\operatorname{eV}%
$ and the mass of $\nu_{\mu}\ $is estimated to $8.4\times10^{-3}%
\operatorname{eV}%
$ for \textit{LMA} and $2.5\times10^{-3}%
\operatorname{eV}%
$ for \textit{SMA}. The graviton rest mass has been estimated by this approach
$m_{g}\sim\hbar H/c^{2}\approx1.5\times10^{-33}%
\operatorname{eV}%
$. Besides, the last value has been obtained independently by dimensional
analysis by means of the fundamental constants $c$, $\hbar\ $and $H$.

It has been found that the rest energy of \textit{LFMSP}, acted upon by a
particular interaction, is close to Breit-Wigner's energy width of the
shortest living state, decaying by the respective interaction. It has been
shown that the mass formula for free massive stable particles $m_{i\min}%
=\hbar/(c^{2}\tau_{i\min})\ $probably involves the Uncertainty Principle.

\bigskip

\end{document}